TITLE: **Seasonal variability in a 1600 year-long ice core chemical record, Pamir Mountains, Central Asia**

AUTHORS: Rodda, Charles*,1,2, Mayewski, Paul,1,2, Kurbatov, Andrei,1,2, Aizen, Elena3, Aizen, Vladimir3, Korotkikh, Elena1, Takeuchi, Nozomu4, Fujita, Koji5, Kawamura, Kenji6, Tsushima, Akane7
1. Climate Change Institute, Orono Maine, USA.
2. School of Earth and Climate Sciences, University of Maine, Orono Maine, USA.
3. Department of Geography, University of Idaho, Moscow, Idaho, USA.
4. Department of Earth Sciences, Chiba University, Chiba, Japan.
5. Graduate School of Environmental Studies, Nagoya University, Nagoya, Japan.
6. National Institute of Polar Research, Tachikawa, Japan.
7. Meteorological Research Institute, Tsukuba, Japan.

ABSTRACT:

Targeted ultra-high resolution Laser Ablation Inductively Coupled Plasma Mass Spectrometry (LA-ICP-MS) analysis of an ice core drilled in the Pamir Mountains of Central Asia (CA) records changes in seasonal patterns of atmospheric circulation and moisture delivery to CA over the past 1600 years. Fe, Ca, and Mg signals in the LNC3 Pamir ice core reflect variable delivery of these elements during the pre-Medieval Warm Period (pre-MWP), Medieval Warm Period (MWP), and during the onset of the Little Ice Age (LIA) in CA. During the pre-MWP, hiatuses between dust "seasons" are observed due to a lengthened accumulation season, and dust delivery limited to a short period at the end of the accumulation season when modern summer-style northerly airflow occurred. During the MWP, dust delivery to CA decreased, though Fe-intensity in the core indicates episodic northerly air incursion. By the onset of the LIA, dust availability increased due to regional drying and strengthening winds, and strong westerlies blocked meridional flow reducing Fe-rich dust input from the north. CA has experienced recent warming, desiccation, weakening winds, and glacier loss. Major climate circulation drivers (Westerlies, Icelandic Low, Siberian High) continue to display LIA-style patterns, and annual temperatures are approaching those of the MWP. While none of the preceding climate periods in CA history can serve as an absolute wholesale proxy for conditions that are likely to prevail in the near future continued Northern Hemisphere westerlies weakening in a changing climate may enhance meridional flow and bring more moisture into the Aral Sea catchment from the Arctic.

INTRODUCTION:

Central Asia (CA) is one of the most arid regions on Earth, and is particularly sensitive to changes in seasonal moisture delivery (Narisma, et al., 2007). Over at least the last two millennia, climate proxy records (see reviews in Chen, et al., 2008; Rodda et al., 2019a) show that moisture delivery to CA has varied widely. The Pamir Mountain

Range in CA is home to dozens of glaciers (Lembrecht et al, 2014) which serve as critical repositories for precipitation, feeding meltwater to multiple major river systems, including the Amu-Darya, the most melt water-dependent major river system on Earth (Viviroli et al., 2004). Indeed, roughly 80% of Central Asia's surface water is sourced from snow and ice melt in the Pamir Mountains (Froebrich and Kayumov, 2004; Kaser et al., 2010; Rodda et al., 2019a) so variation in the amount of moisture delivered to CA glaciers fundamentally drives moisture availability downstream, for both natural biotic communities, and the ~72 million inhabitants of the region (IMF, 2018).

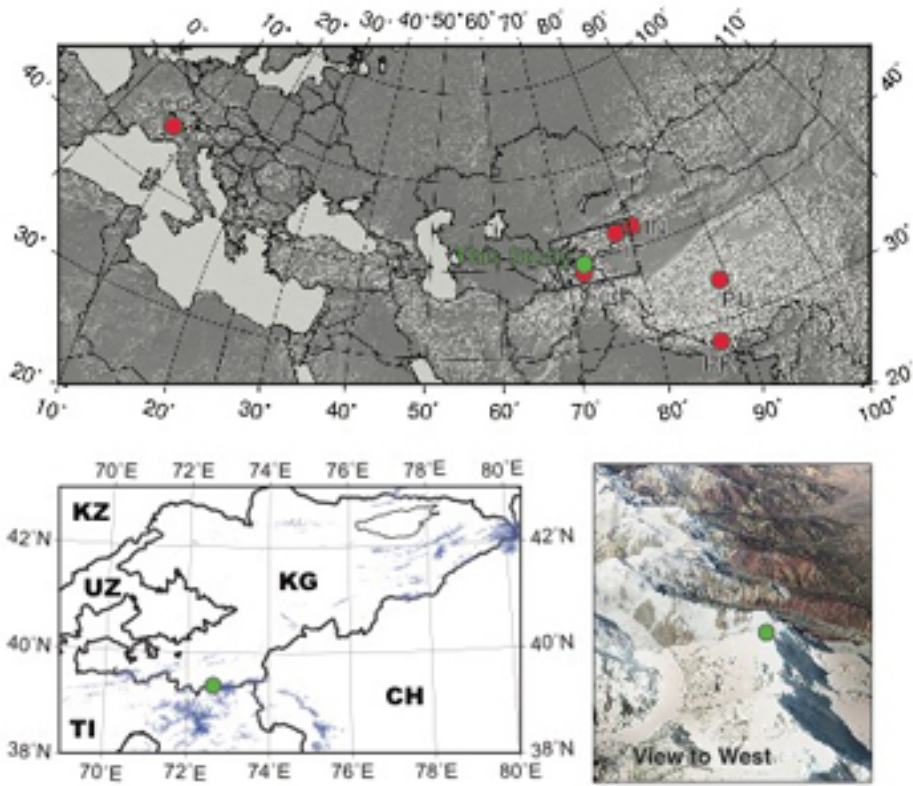

Figure 1.
Research site location map, including the location of the LNC3 ice core referred to later in the text marked by a green dot in all panels. Panel a. DEM of Central Asia (CA), upwind portions of Eurasia and close areas of Eastern Asia. Panel b. Glacierized portions of CA, with glaciers indicated in blue. Country abbreviations used in associated manuscripts are indicated – KZ: Kazakhstan; UZ: Uzbekistan; TI: Tajikistan; TX: Turkmenistan; KG: Kyrgyzstan; AF: Afghanistan; CH: China. Panel c. Google Earth 3D image of drill site. The aerial photograph shows a view of the LNC3 drill site on a dome in the KG/TI border region in the glacierized Pamir Mountains. Image view is from immediately East of the drill site, looking West.

Kreutz, et al. (2003) suggest that moisture delivery from the Caspian/Aral Sea region does not penetrate beyond the high mountains of the Pamir, Tien Shan and Hindu Kush, into China or Northern India. On both seasonal and century-scales, climate patterns are asynchronous between the westerlies-dominated area (Chen et al, 2008; Grigholm, et al., 2015) and the Caspian Sea and the Pamir Range, and those regions farther east, dominated by the Indian and Southeast Asian Monsoons (Chen, et al., 2008). Due to the spatial (Kreutz, et al., 2003; Unger-Shayesteh, et al., 2013) and temporal (Chen et al., 2008; Unger-Shayesteh, et al., 2013) variability in atmospheric circulation and moisture delivery across Eurasia, characterizations of climate change in CA must rely on long term records within the westerlies-dominated Aral Sea region. In a recent study, Rodda et al. (2019a) note that over the length of the Global Historical Climatology Network (GHCN) weather station record for CA (1890s-present), there exists a narrow and shrinking spring (MAM) season in which spring precipitation begins, and when temperatures in the glacierized highlands remain cold enough for precipitation to accumulate as snow on glaciers and snowfields. Small shifts in the seasonal timing or patterns of temperature and precipitation threaten this delicate balance. That study also documented that in the last decade the glacierized highlands of CA have experienced a mean decrease of 2-5cm/a in precipitation during the critical May cold/wet season and +1*C of warming, compared to the previous 27 years of satellite data, which itself represents an already warming baseline, compared to preceding decades. Taken together, these patterns are consistent with the significant CA glacier retreat noted by Aizen, et al. (2016) between 1970 and 2010.

An overall wet-spring/dry-summer/moderate-winter seasonal pattern holds for all countries besides Kazakhstan (which is relatively dry throughout the year). The wet season begins in February or March; earliest in those countries with more high-elevation regions (Tajikistan, Afghanistan, Kyrgyzstan). As discussed in Rodda, et al. (2019a) the GHCN data for Tajikistan does not reflect sub-freezing temperatures in the highlands during DJF and MAM due to the non-representative inclusion of primarily low-elevation sites in the GHCN record.

Seasonal mean reanalysis maps illustrate seasonal patterns of atmospheric circulation (figures 2 and 3) and atmospheric moisture availability (figure 4) for CA. Wind patterns at 10m above the surface (figure 2) show little spatial coherence, and low velocity overall from the autumn (SON) through the spring season (MAM). Vectors indicate air motion in nearly all directions, reflecting local topographic control over regional circulation patterns during these months. The summer months (JJA) display a minor exception to this rule as northerly surface winds strengthen and flow across Kazakhstan into the Aral Sea drainage. This northerly air mass incursion at the surface is accompanied by weakening of the westerlies aloft (figure 3).

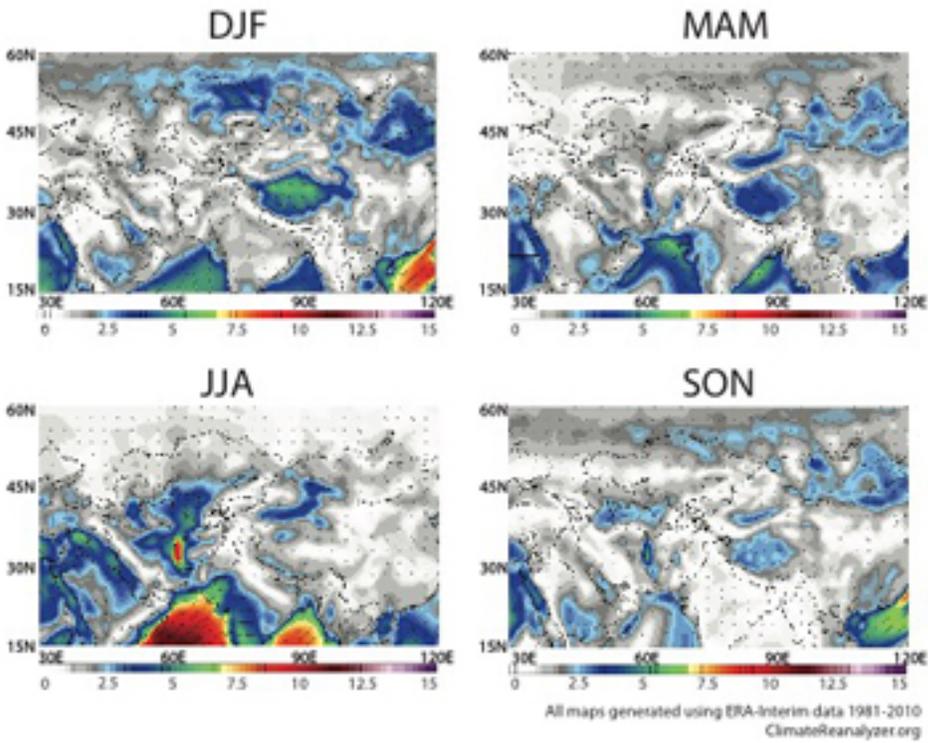

Figure 2. Seasonal 10m mean winds in CA and the surrounding area for the period 1981-2010. Maps use ERA-Interim data, (compiled in Climate Reanalyzer, 2019).

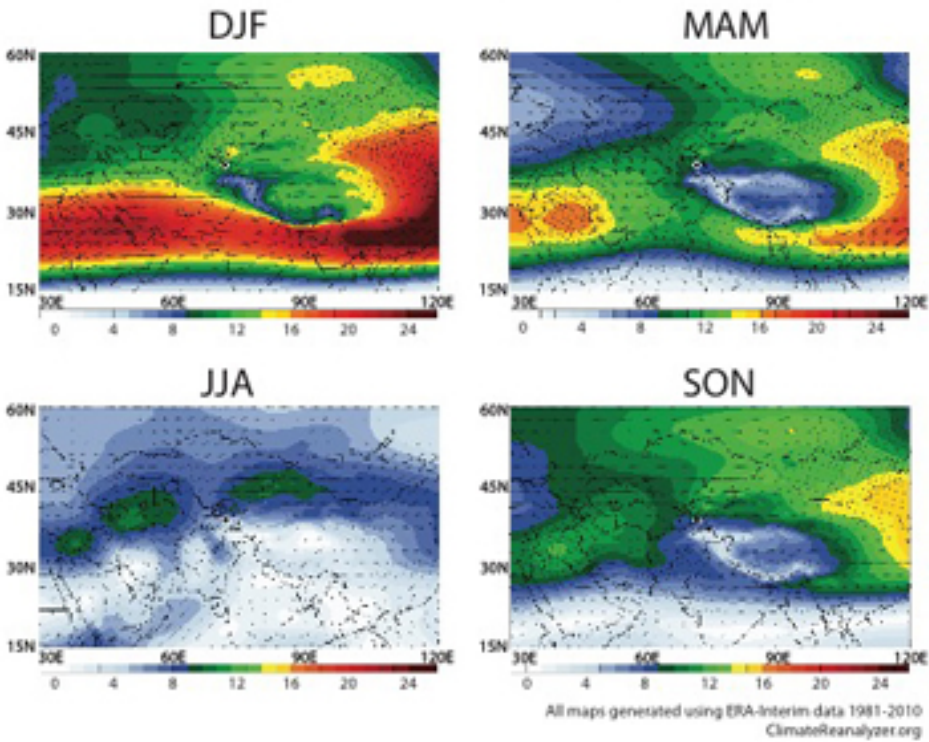

Figure 3. Seasonal mean winds at 500 hPa pressure level over CA and the surrounding area for the period 1981-2010. Maps use ERA-Interim data, (compiled in Climate Reanalyzer, 2019).

Over much of Asia, the summer weakening of the westerlies is coincident with the moist phase of the Indian and Southeast Asian monsoon. The summer incursion of moisture from the Indian Ocean is readily apparent in a reanalysis map of atmospheric precipitable water (figure 4, JJA panel). That Indian Monsoon moisture does not currently penetrate into CA, however, increased summer northerly flow brings additional moisture from the arctic across Kazakhstan, into CA.

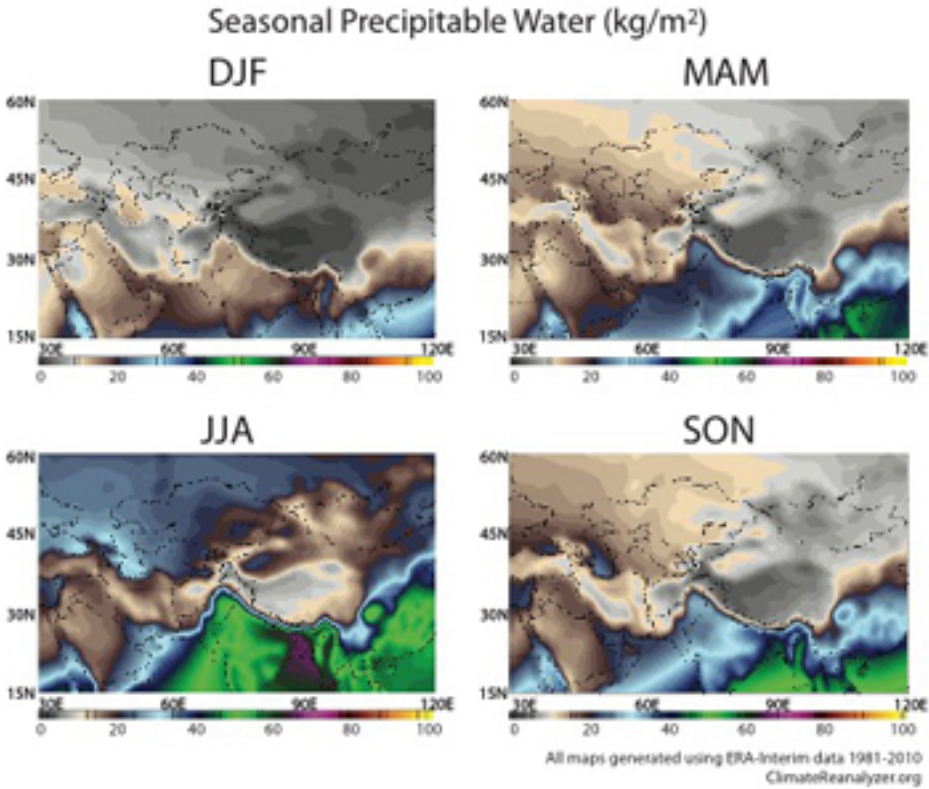

Figure 4. Seasonal precipitable moisture over CA and the surrounding area for the period 1981-2010. Maps use ERA-Interim data, (compiled in Climate Reanalyzer, 2019).

Existing annually-resolved ice core records from CA (Fedchenko Glacier (Aizen, et al., 2009), It-Tish Ice Cap (Thompson, et al., 1997), Grigorieva Ice Cap (Thompson, et al., 1997; Okamoto et al., 2011) and Inilchek Glacier (Aizen et al., 2006)) cover roughly the last 100 years, or less. 20[th] century CA climate is also well documented in weather station records (Lawrimore et al., 2011). Ages in the 12,650 year-long Grigorieva Ice Cap (Takeuchi, et al., 2014) record are constrained by a basal soil carbon age, and discrete dust layers >5000 years old; but the age scale lacks constraint between the onset of the Medieval Warm Period (MWP), and the present, including the transition into the Little Ice Age (LIA).

An ice core recently collected in the Pamir Mountains includes sub-annual-scale dust-element records from periods dating to ~400 C.E. (Rodda et al., 2019b). In this paper using ultra-high resolution (120 micron) laser ablation inductively-coupled plasma mass spectrometry (LA-ICP-MS) dust element records (Ca, Mg, Fe) from that core we examine the distinctly different atmospheric circulation/temperature/precipitation patterns of the Pre-MWP at ~475 C.E. and the Medieval MWP at ~1025 C.E., both with warmer-than-present conditions, and near the onset of the LIA at ~1425 C.E., with generally cold/dry/windy conditions. The sections under investigation were chosen based on their occurrence during each of the climate regimes noted above balanced against the availability of sufficient quality ice in the LNC3 core. Many sections of this ice core were fractured making them unusable for the laser sampling technology.

METHODS:

The Pamir International High Elevation Geophysical Expedition/International Geoscience Programme (HEIGE IGCP Project 650) expedition team travelled to the Kyrgyzstan-Tajikistan border region (39.39N, 72.63E, 5478m a.s.l., figure 1) in September 2016, and recovered an ice core that terminated in bedrock (the LNC3 core). Description of the collection, transport and initial processing of the core, including rationale for the timescale developed for the LNC3 record can be found in Rodda et al. (2019b).

The LA-ICP-MS data for this study utilized the apparatus and methods originally developed by Sneed et al. (2015) in the Keck Laser Ice Facility LA-ICP-MS system at CCI. Intact core samples are placed into a sub-0°C cryo-chamber, and maintained in a frozen state while non-destructive sampling occurs. The sampling system uses a focused laser to ablate a nearly-microscopic trough along a freshly-scraped surface. Sample resolution from the system is ~50,000 per m of core depth. LA-ICP-MS sampled elements: Al, Ca, Fe, Mg, Cu, and Pb.

For this study, a lengthwise split of the core was cut, using specially adapted band saws in the ice core cold (-20°C) lab at Helmholtz Centre for Polar and Marine Research, Alfred Wegener Institute (AWI, Germany), and transferred to the Climate Change Institute (CCI) in Orono, Maine, USA for ultra-high resolution laser-ablation inductively-coupled-plasma mass spectrometry (LA-ICP-MS) sampling. Following development of the age model for the LNC3 ice core, sections of clear, crack-free ice dating to the Pre-MWP, MWP, and LIA were selected for LA-ICP-MS analysis. On each sample, a fresh surface orthogonal to annual layers was exposed by hand-scraping with clean ceramic blades in the CCI ultra-cold (-20*C) ice core clean lab to ensure that an uncontaminated, unaltered portion of the core was sampled.

RESULTS/DISCUSSION:

Mg, Ca, and Fe are environmental indicators of dust input in ice cores (Shichang et al., 2000; Kang et al., 2002; Ruth et al., 2008; Dong et al., 2014), so we routinely utilize these elements to define annual layers in LA-ICP-MS ice core records. For each of the three climate periods in the CA record, we measured intensity of these elements at ultra-high resolution (120-micron sampling) (Sneed et al., 2015). Based on the annual layer thicknesses observed, this results in an annual sampling resolution for each period as follows - Pre-MWP: ~100 samples/a; MWP: ~245 samples/a; LIA: ~500 samples/a. At all depths, the resolution is sufficient to distinctly record individual storm events.

To ensure that sample resolution does not drive our interpretation, we smoothed shallow (higher temporal resolution) data sets to an interval that reflects the relatively lower temporal resolution of the deepest samples (figure 5). The smoothed dataset shows the same number of precipitation events as are recorded in the full dataset, ensuring that sampling resolution is not introducing erroneous patterns.

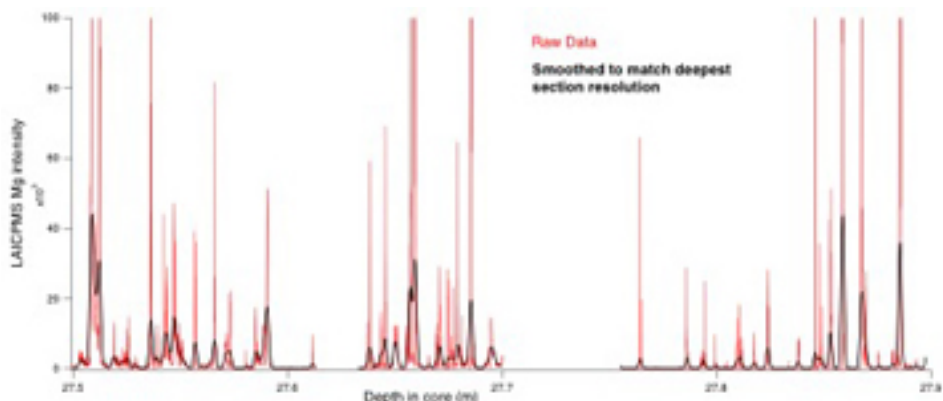

Figure 5. Full-resolution (~500 samples/a) and re-sampled (~100 samples/a) LIA-period Fe records. The smoothed curve reflects the same temporal resolution that deeper, more compressed layers in the LNC3 core have. Both the raw and the smoothed curves show the same number and similar intensity of peaks, demonstrating that dust signal intensity and peak intervals are not resolution-induced phenomena.

**The Pre-MWP (example ca. 475 C.E.) LAICPMS record:**

During the Pre-MWP, all three dust element records display single, coeval annual peaks (figure 6, top panel). Minor details of the individual annual peak shapes, such as minor asymmetries and secondary accessory peaks are reflected across all three elements. The similarity of the records indicates that the circulation pattern delivering all of these elements is the same. Compared to later in the record (figure 6, lower panels), the high intensity of Fe relative to Ca and Mg in the Pre-MWP suggests that the annual dust peak is delivered by storms that track over an area with abundant Fe-rich dust.

The single, relatively smooth annual peaks displayed during the Pre-MWP indicate that atmospheric circulation patterns that deliver dust to the LNC3 core site set-up during the accumulation period for the year, remain relatively steady at their peak intensity, and then that system breaks down. Between annual peaks, a background intensity several orders of magnitude lower than the annual peak exists for most of the thickness of each annual layer. This quiescence is not displayed during the other periods sampled. The low dust signal indicates that conditions either local to the core site or in the dominant moisture source region precluded the incorporation of dust but not moisture into the airmass, that the winds were too low to incorporate dust, or that the dust was rained-out before reaching the glacier.

As noted in Rodda et al. (2019b), annual layer thicknesses predicted by glacier flow models and those observed in the LNC3 core are in excellent agreement throughout most of the core's length. In this section of the core, however, annual layers are nearly twice as thick as flow modeling would suggest. This reflects either increased moisture delivery to the region during this time, relative to other periods in CA, or conditions more favorable for annual snowfall to transition into glacier ice, rather than melting during the intervening summer season.

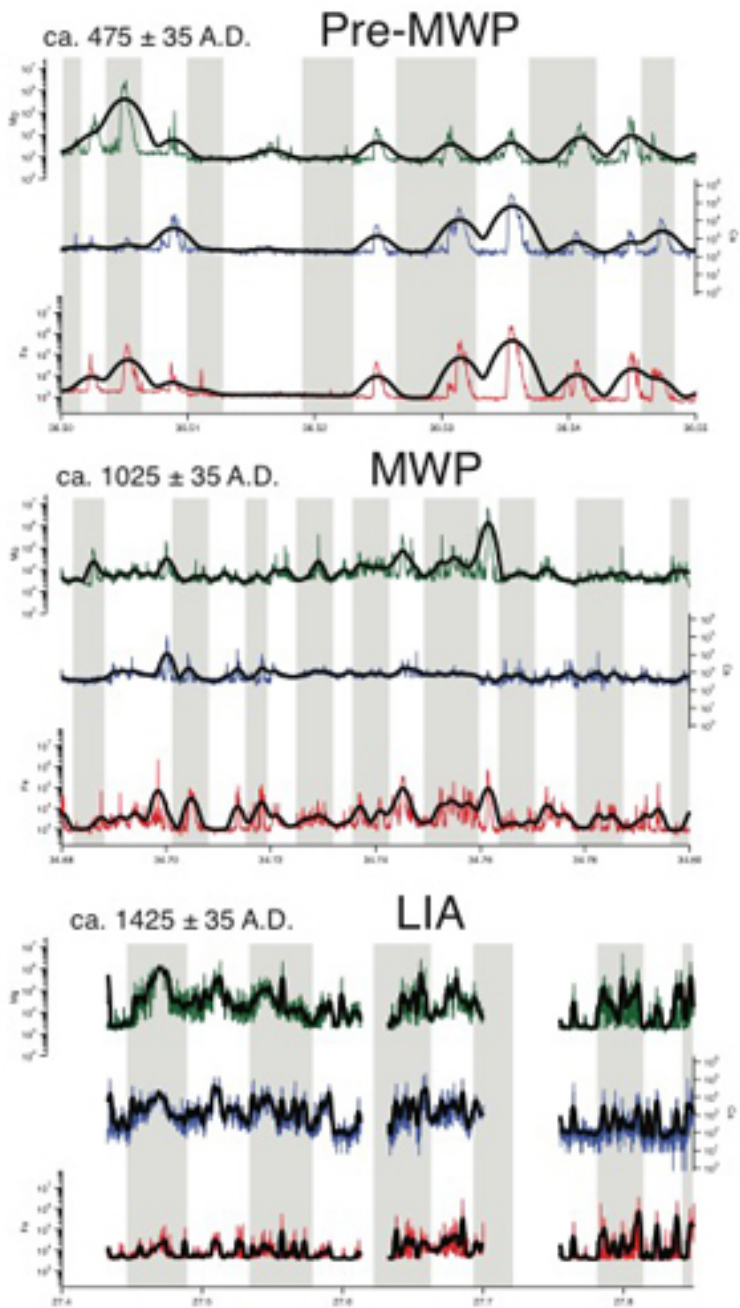

Figure 6. Top panel: approximately 12 annual dust peaks in the LA-ICP-MS signal from the Pre-MWP (Ca. 475 C.E.) in CA. Middle panel: approximately 16 annual dust peaks in the LA-ICP-MS signal from the Pre-MWP (Ca. 1025 C.E.) in CA. Bottom panel: approximately 10 annual dust peaks in the LA-ICP-MS signal from the Pre-MWP (Ca. 1025 C.E.) in CA. In all panels, elemental curves are Mg (red), Ca (blue), and Fe (green). Alternating white and grey background shading marks individual years.

**Existing Pre-MWP Paleo records:**

Stable isotopic enrichment in the Altai indicates a prolonged period of warm wet seasons lasting from ~2.9k to 950 yBP, (Aizen, et al., 2014). A warm period is also indicated by growth of Juniperus turkistanica on a proglacial terrace in the Pamir-Alay, beginning ~ 2400 yBP ending when trees were buried by a glacial advance at 1545 yBP (Narama, 2002). From ~2.6k yBP until ~1.6k yBP, an Aral Sea moisture index, constructed from shore terrace records, historical documents, calculation of sedimentation rates and CaCO3 variations, reflects continual drying, attaining maximum-dry conditions at ~1.6k BP (Chen et al., 2008). High concentrations of high-salinity dinoflagellate cyst assemblages and gypsum deposition in the Aral Sea Basin indicate low lake levels, with prolonged periods of late summer evaporation, beginning ~2.1k BP and lasting until ~1575 BP (Sorrel et al., 2006). Aral Sea salinity increases, indicated by an increase in concentration of dinoflagellates with high salinity affinity, during the period ~2.1k-1575 BP, occurring during several short (<100 y) periods likely resulting from decreased meltwater supply to the Amu Darya and Syr Darya associated with low spring and early summer temperatures in the glaciated upper reaches of these rivers' catchments (Sorrel, et al., 2006). Cold spring/early-summer temperatures are consistent with expansion of glaciers in the Pamir Mountains during this period (Zech et al., 2000; Narama, 2002). Total aeolian dust delivery to the Aral Sea during the period ~2k-1650 BP was low, and characterized by low-Ti content (titanium serves as an indicator of wind-borne sediments in that basin) and fine grain sizes, consistent with the lowest spring wind intensity during the past 2000 years (Xiangtong, et al., 2011). From ~2k - ~1575 BP, Sorrel et al. (2006) also note abundant freshwater algae and nutrient-dependent dinoflagellates in Aral Sea sediments indicative of periodic, abundant freshwater discharge into the basin. It is not clear whether the source of that enhanced freshwater input was the Amu Darya and Syr Darya or more local sources (Sorrel et al., 2006; Aleshinskaya et al., 1996). From ~2.1 yBP until ~1575 yBP the coincidence of cool-water spring-blooming species with high-salinity summer-blooming species points to a higher seasonal temperature difference, with warm summers and cold winters (Sorrel et al., 2006).

**Pre-MWP Discussion**

Since the Aral Sea Basin, up-wind of the Pamirs, was drying during the Pre-MWP, it should have been a good and potentially improving source of dust for the Pamir ice core site, and likely precludes significant increases in moisture being the (sole) cause of apparently thick annual layers. The long annual low-dust period in the LA-ICPMS record, therefore, reflects the low-wind conditions also recorded in Aral Sea sediment cores. Cold spring temperatures would likely have led to a lengthened accumulation season, allowing annual layers to be preserved and therefore relatively thicker. Local sources of dust to the glacier would have been minimized, with cold spring temperatures leading to increased persistence of seasonal snowfields at their margins, and plants colonizing pro-glacial terraces,

**The MWP (example ca. 1025 C.E.) LAICPMS record:**

During the MWP, the annual dust element hiatuses observed in the Pre-MWP LA-ICP-MS record are absent, or diminished to very brief periods preceding or following the peaks that take up the majority of the record (figure 6, middle panel). Whatever circulation patterns or local phenomena allowed for low-dust moisture to access the LNC3 site at 475 C.E. was no longer available by 1025 C.E. Relative to the Pre-MWP, annual peaks are diminished in total intensity, especially in the Ca record. Most years in the MWP LA-ICP-MS record still display a single primary peak in Ca and Mg, and those peaks are generally in-phase, though commonly are of dissimilar intensity (i.e., see 34.73m depth). The Fe record from the MWP shows more variability than the Mg and Ca records. The Fe record has a greater number of peaks per year, and in some years, the timing and relative intensity of the overall peaks are not consistent with the other elements (i.e. laser Fe at 34.70m depth). Dissimilar timing and intensity of the major annual peak relative to Mg and Ca suggests that Fe-rich dust may be sourced from a region that is different from that which brings in much of the other dust in the core. Major iron ore deposits in Europe are clustered in Western Europe from the Atlantic coast extending roughly to the middle of Mediterranean Sea (Dottin and Gabert, 1990). Major Fe-bearing loess deposits extend from the middle of the Mediterranean, through the Black and Caspian Sea regions and nearly to the CA field area (Fink et al., 1977).

**Existing MWP Paleo records:**

During the MWP, weakening of the Eastern Asian summer monsoon and reorganization of the westerly wind system likely resulted in a change in the dominant source region for moisture delivered to the Altai, with a decrease in Atlantic moisture and a corresponding increase in the contribution of re-evaporated precipitation from the Caspian and Aral Sea region (Aizen et al., 2014; Bradley, 2000). This is consistent with larger-scale circulation patterns noted by Meeker and Mayewski (2002), using the Greenland Ice Sheet Two (GISP2) ice core glaciochemically reconstructed atmospheric circulation proxies that found for the MWP generally weaker mean values for the Icelandic Low during DJF and the Siberian High during MAM.

Peak tree ring widths observed in CA occurred at ca.1000yBP; 81% of tree ring widths in the period 1099-900yBP are higher than the 1300+ year average for the region and almost no trees show negative growth trends during that period (Esper, et al., 2002).

Lush tree growth occurred on thick soils developed during the same period on terraces in front of the Raigorodskogo Glacier, watered by heavy meltwater discharge (Narama, 2002).

**MWP Discussion**:
Since the Westerlies were diminished in strength during this period, it is unlikely that relative increases in Fe intensity and deposition events in the MWP indicate stronger circulation between a Western Europe source region and CA. Diminished peak intensities in Mg and Ca are also consistent with weaker winds.

Fe input to the core remains at or above Pre-MWP levels during the MWP, and displays many more distinct deposition events throughout each year. Mg displays coeval peaks with Fe in some seasons, and with Mg in others, and all three elements seem to be less well temporally correlated. It is likely that air masses reaching the LNC3 site experienced less coherent up-stream rain-outs of their dust load during the MWP, and storms at that time delivered more irregular bursts of dust, from heterogeneous, (hyper-)localized source regions within the Aral Sea catchment. In modern climate reanalyses, weaker westerlies during the JJA summer season allow stronger meridional flow across Kazakhstan into the Aral Sea drainage. If the MWP weakening of the westerlies led to similarly enhanced meridional northerly flow, Fe peaks likely relate to increased incursion of airmasses that traveled across Fe-rich areas. A concentration of 19$^{th}$ and 20$^{th}$ century iron mines, landscape-scale red-sand filled valleys and massive Fe-oxide stained outcrops to the north of the Pamir LNC3 core site are likely source regions for Fe-rich dust (figure 7).

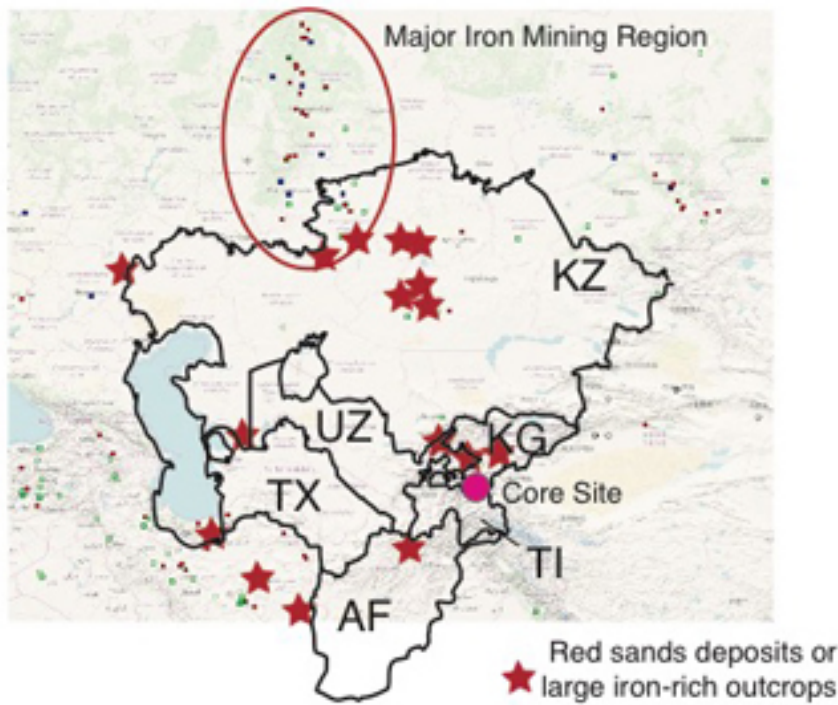

Figure 7. Potential sources of Fe-rich dust in CA. Red stars indicate landscape-scale red sand deposits and apparent FeOx-stained outcrops observed in aerial imagery (Google Earth Pro, 2019). Colored dots indicate iron mines identified in the USGS Mineral Resources database (Mason and Arndt, 2019).

**The LIA-onset (example ca. 1425 C.E.) LAICPMS record:**

During the portion of the LNC3 LA-ICP-MS record dating to the onset of the LIA, total dust intensity is much higher than in earlier portions of the record (figure 6, bottom panel). The Pre-MWP no-/low-dust hiatus is totally absent from this portion of the record. Fe peaks are similar in maximum intensity to those from deeper portions of the record, suggesting that the Fe source area and delivery remained similar throughout the transitions between climate periods, as major changes in circulation and precipitation patterns occurred throughout the region. Both the average intensity and number of delivery events per year observed in the Mg and Ca records are much higher during this period than earlier portions of the record, and co-occurrence between the two signals is once again strong. That relationship suggests a strong, yet variable, circulation pattern is controlling bulk dust delivery to the glacier site, and the dissimilarity of the Fe record to the other two elements indicates that the Fe source site was able to deliver Fe-rich dust to the region only sporadically. If the Fe source was in fact the Fe-rich region to the north, strengthening westerlies during the LIA would likely have diminished the incursion of northerly air masses, though highly variable inter- and intra-annular conditions in CA observed during the LIA (Rodda et al., 2019b) would explain irregular strong northerly flow.

**Existing LIA-onset Paleo records:**

From the end of the MWP to the top of the Altai Belukha ice core, stable water isotope records indicate increasingly mixed and re-cycled moisture sources, with distinct provenances – Pacific, North Atlantic/Arctic, Atlantic, and continental re-evaporation of precipitation coming from the Mediterranean, Black and Aral Seas to the West, (Aizen et al., 2005, 2006, 2014).

During the LIA onset, salinity was low in the Aral Sea Basin, and likely decreased further during the first stages of the LIA, indicating continued freshwater inputs to the basin (Sorrel, et al., 2006). During two short periods (ca., ~500 and ~400-350yBP) the basin aridified, evidenced by the return of more brackish dinoflagellates and the deposition of a gypsum layer; these moderately brackish conditions persisted until the end of the LIA (Sorrel et al., 2006). These two arid periods coincide with the coldest decades in the last millennium, based on mean reconstructions for the Northern Hemisphere (Bradley et al., 2000). Interestingly, the only de-coupling of growth patterns in trees from the Tien Shan region in Kyrgyzstan and the Karakorum in Pakistan occurs between ~625 and ~400yBP (Esper et al., 2002), suggesting regional imbalance between temperature-limited and precipitation-limited growth pattern shifts or highly localized micro-climatic transitions into LIA conditions. Esper et al. (2002) also note that the worst tree growing conditions measured in the last 1,373 years were attained ~500yBP, and lasted for more than a century. Taft et al. (2014) interpret the gastropod and ostracod isotopic patterns in Lake Karakul as indicating cool and dry conditions from ~500 – 300yBP, though a slight warming (while still dry) may have occurred ~400yBP (Mischke et al., 2010).

The most abrupt increase in Ti-content and size fraction of Aral Sea sediments observed during the last 2000 years occurs a ~600 to ~570yBP (Xiangtong, et al., 2011). It is interpreted as a sudden increase in wind strength at the onset of the LIA. The Aral Sea moisture index drops from wet to dry at ~600 BP, and remains there until the end of the record (Chen, et al., 2008). Total aeolian dust delivery to the Aral Sea remained high throughout the LIA, with a peak at 390-380 BP (Xiangtong, et al., 2011). Consistent with the foregoing, Meeker and Mayewski (2002) find that ~600-580yBP there is an abrupt shift in nssK and ssNa concentrations in the GISP2 ice core, signaling a transition from weak winter (DJF) Icelandic Lows to more intense circulation over the North Atlantic and a simultaneous transition from weak spring (MAM) Siberian Highs to a period of strengthened circulation.

**LIA-onset Discussion:**

Existing published LIA climate proxy records indicate cooling, and intensification of winds during the onset of the LIA. Moreover, many of the records indicate more temporal variability and decreased regional coherence of climate signals. The LNC3 core also records a significant increase in total dust delivery during the onset of the LIA reflecting increased winds. Because the intensity of Fe in the core does not demonstrate a coeval increase, we assume that its source is the Fe-rich region to the north, and increased availability of Fe-rich dust in that area resulted in increased Fe-intensity in the LNC3 record only when meridional flow was possible during the summer (JJA) season or during storm events that brought northerly air masses into the Aral Sea drainage.

CONCLUSIONS:

Fe, Ca, and Mg signals in the LNC3 Pamir ice core reflect changes in atmospheric circulation. Ultra-high resolution LA-ICP-MS analysis of pre-MWP, MWP, and onset of the LIA in CA reveal differences in the relative intensity and coherence of these three sets of seasonal scale signals illuminating regional and local scale variations in atmospheric circulation.

An apparently local high-Fe dust source to the north showed similar intensity and timing to Ca and Mg signals during the Pre-MWP, at a time when long hiatuses between dust "seasons" are preserved in the record. The region was wetter than present over this period, decreasing dust in the atmosphere. Cool spring weather at that time likely lengthened the accumulation season, and dust delivery to the core site may have been limited to a short period at the end of the accumulation season when modern summer (JJA)-style northerly airflow occurred.

During the MWP, when warmer-than-present conditions and weak general circulation patterns prevailed in CA, Fe input to the core remained similar to the Pre-MWP in intensity, though it became more episodic. The phenomena that precluded dust delivery during part of the year during the pre-MWP disappeared, and Fe was delivered with northerly storms that also brought in variable amounts of Mg and Ca, depending on the precise storm track.

By the onset of the LIA, strengthening Westerlies delivered more dust to the LNC3 core. Again, Fe intensity remained largely unchanged, and unconnected to the gross dust input. Dust availability increased due to regional drying and strengthening winds, and strong westerlies blocked meridional flow reducing Fe-rich dust input from the north. This brings into sharp clarity the fact that, in the Pamir Mountains, very local phenomena can exert strong controls on glacierized regions.

CA has experienced warming, desiccation, weakening winds and decreasing glacier and snowfield area in recent decades. Though many of the major climate circulation drivers (Westerlies, Icelandic Low, Siberian High) continue to display LIA-style patterns, annual temperatures are approaching those of the MWP. Thus far, the climate periods we have examined in CA history do not serve as an absolute proxy for conditions that are likely to prevail in the coming decades. However, if the Northern Hemisphere westerlies continue to weaken, enhanced meridional flow will bring more moisture into the Aral Sea catchment from the Arctic, essentially increasing JJA-type atmospheric flow. Whether or not that moisture can be captured in Pamir Range glaciers warrants very careful tracking, and whether or not the input timing of this moisture is during a season when it can be preserved as a water tower source is the critical question since these glaciers provide critical water resources to human and non-human communities alike.

ACKNOWLEDGEMENTS:

This research was funded by NSF grant #1401899. It benefitted greatly from the involvement of Frank Wilhelms, York Schlomann, and Pascal Bohleber (Helmholtz Centre for Polar and Marine Research, Alfred Wegener Institute), Azamat Osmonov (Central Asian Institute for Applied Geosciences), Nargiza Shaidyldaeva (Xinjiang Institute of Ecology and Geography), and Sharon Sneed, Nicky Spaulding, and Heather Clifford (Climate Change Institute, University of Maine).